# Understanding Class-level Testability through Dynamic Analysis

Amjed Tahir[1, *], Stephen G. MacDonell[2] and Jim Buchan[1]

[1]*Software Engineering Research Laboratory, Auckland University of Technology, Auckland, New Zealand*
[2]*Department of Information Science, University of Otago, Dunedin, New Zealand*
*{amjed.tahir, jim.buchan}@aut.ac.nz, stephen.macdonell@otago.ac.nz*

Keywords:     Software Testability, Dynamic Metrics, Dynamic Analysis, Unit Testing, Software Understanding.

Abstract:     It is generally acknowledged that software testing is both challenging and time-consuming. Understanding the factors that may positively or negatively affect testing effort will point to possibilities for reducing this effort. Consequently there is a significant body of research that has investigated relationships between static code properties and testability. The work reported in this paper complements this body of research by providing an empirical evaluation of the degree of association between runtime properties and class-level testability in object-oriented (OO) systems. The motivation for the use of dynamic code properties comes from the success of such metrics in providing a more complete insight into the multiple dimensions of software quality. In particular, we investigate the potential relationships between the runtime characteristics of production code, represented by Dynamic Coupling and Key Classes, and internal class-level testability. Testability of a class is considered here at the level of unit tests and two different measures are used to characterise those unit tests. The selected measures relate to test scope and structure: one is intended to measure the unit test size, represented by test lines of code, and the other is designed to reflect the intended design, represented by the number of test cases. In this research we found that Dynamic Coupling and Key Classes have significant correlations with class-level testability measures. We therefore suggest that these properties could be used as indicators of class-level testability. These results enhance our current knowledge and should help researchers in the area to build on previous results regarding factors believed to be related to testability and testing. Our results should also benefit practitioners in future class testability planning and maintenance activities.

## 1 INTRODUCTION

Software testing is a core software engineering activity. Although software systems have been growing larger and more complex for some time, testing resources, by comparison, have remained limited or constrained (Mouchawrab et al., 2005). Software testing activities can also be costly, requiring significant time and effort in both planning and execution, and yet they are often unpredictable in terms of their effectiveness (Bertolino, 2007). Understanding and reducing testing effort have therefore been enduring fundamental goals for both academic and industrial research.

---

* A. Tahir is now with the Department of Information Science, University of Otago, New Zealand

The notion that a software product has properties that are related to the effort needed to validate that product is commonly referred to as the 'testability' of that product (ISO, 2001). In fact, this term can be traced back to 1994, when Binder (1994) coined the phrase "Design for Testability" to describe software construction that considers testability from the early stages of the development. The core expectation is that software components with a high degree of testability are easier to test and consequently will be more effectively tested, raising the software quality as compared to software that has lower testability. Improving software testability should help to reduce testing cost, effort, and demand for resources. Traon and Robach (1995) noted that if components are difficult to test, then the size of the test cases designed to test those components, and the required testing effort, will necessarily be larger. Components with poor testability are also more expensive to repair when problems are detected late in the development process. In contrast, components and software with good testability can dramatically increase the quality of the software as well as reduce the cost of testing (Gao et al., 2003).

While clearly a desirable trait, testability has been recognised as being an elusive concept, and its measurement and evaluation are acknowledged to be challenging endeavours (Mouchawrab et al., 2005). In spite of the ISO definition, or perhaps because of its rather broad meaning, multiple views have been adopted when authors have considered software testability. Researchers have therefore identified numerous factors that (may) have an impact on the testability of software. For instance, software testability is said to be affected by the extent of the required validation, the process and tools used, and the representation of the requirements, among other factors (Bruntink and van Deursen, 2006). Given their various foundations it is challenging to form a complete and consistent view on all the potential factors that may affect testability and the degree to which these factors are present and influential under different testing contexts. Several are considered here to provide an initial overview of the breadth of factors of potential influence on testability.

A substantial body of work has addressed a diversity of design and code characteristics that can affect the testability of a software product. For example, the relationships been internal class properties in OO systems and characteristics of the corresponding unit tests have been investigated in several previous studies e.g., Bruntink and van Deursen (2006), Badri et al., (2011). In these studies, several OO design metrics (drawn mainly from the C&K suite (Chidamber and Kemerer, 1994)) have been used to investigate the relationship between class/system structure and test complexity. Some strong and significant relationships between several complexity- and size-related metrics of production code and internal test code properties have been found (Bruntink and van Deursen, 2006).

In their research, Bruntink and van Deursen used only static software measures, and this is the case for all previous work in this area. In this paper we build on the view of Basili et al. (1996) that traditional static software metrics may be necessary but not sufficient for characterising, assessing and predicting the entire quality profile of OO systems, and so we propose the use of dynamic metrics to represent further characteristics. Dynamic metrics are the sub-class of software measures that capture the dynamic behaviour of a software system and have been shown to be related to software quality attributes (Cai, 2008, Gunnalan et al., 2005, Scotto et al., 2006). Consideration of this group of metrics provides a more complete insight into the multiple dimensions of software quality when compared to static metrics alone (Dufour et al., 2003). Dynamic metrics are usually computed based on data collected during program execution (i.e., at runtime) and may be obtained from the execution traces of the code (Gunnalan et al., 2005), although in some cases simulation can be used instead of the actual execution. Therefore they can directly reflect the quality attributes of that program, product or system *in operation*. This paper extends the investigation of software characteristics as factors in code testability by characterising that code using dynamic metrics. A fuller discussion of dynamic metrics and their relative advantages over static metrics is presented in another article (Tahir and MacDonell, 2012).

The rest of the paper is structured as follows. Section 2 provides the research context for this paper by reviewing related work, and confirms the potential of relating dynamic code metrics to testability. Section 3 argues for the suitability of the Dynamic Coupling and Key Classes concepts as appropriate dynamic metrics to characterise the code in relation to testability. These metrics are then used in the design of a suitable set of experiments to test our hypotheses on specific case systems, as described in sections 4 and 5. The results of these experiments are then presented in section 6 and their implications are discussed in section 7. Threats to the study's validity are noted in section 8. Finally, the main conclusions from the study and some thoughts on related future work are presented in section 9.

## 2 RELATED WORK

Several previous works have investigated the relationships between properties of software production code components and properties of their associated test code, with the focus primarily on unit tests. The focus of that work has varied from designing measures for testability and testing effort to assessing the strength of the relationships between them. Given constraints on space, we consider a few typical studies here. Our intent is to be illustrative as opposed to exhaustive, and these studies are representative of the larger body of work in this research domain.

Bruntink and van Deursen (2006) investigated the relationship between several OO metrics and class-level testability for the purpose of planning and estimating later testing activities. The authors found a strong correlation between class-level metrics, such as Number of Methods (NOM), and test level metrics, including the number of test cases and the lines of code per test class. Five different software systems, including one open source system, were traversed during their experiments. However, no evidence of relationships was found between inheritance-related metrics, e.g., Coupling Between Objects (CBO), and the proposed testability metrics. This is likely to be because the test metrics were considered at the class level. These inheritance-related metrics are expected to have a strong correlation with testability at the integration and/or system level, as polymorphism and dynamic binding increase the complexity of a system and the number of required test cases, and contribute to a consequent decrease in testability (Mouchawrab et al., 2005). This suggestion can only be confirmed through evaluation at the object level using dynamic metrics. In a similar study, Badri et al. (2011) investigated the relationship between cohesion and testability using the C&K static Lack of Cohesion metric. They found a significant relationship between this measure of static cohesion and software testability, where testability was measured using the metrics suggested by Bruntink and van Deursen (2006).

In other work related to testability, Arisholm et al. (2004) found significant relationships between Dynamic Coupling measures, especially Dynamic Export Coupling, and change-proneness. Export Coupling appears to be a significant indicator of change-proneness and likely complements existing coupling measures based on static analysis (i.e., when used with size and static coupling measures).

# 3 TESTABILITY CONCEPTS

## 3.1 Dynamic Coupling

In this study Dynamic Coupling has been selected as one of the system characteristics to measure and investigate regarding its relationship to testability. Coupling has been shown in prior work to have a direct impact on the quality of software, and is also related to the software quality characteristics of complexity and maintainability (Offutt et al., 2008, Al Dallal, 2013). It has been shown that, all other things being equal, the greater the coupling level, the greater the complexity and the harder it is to maintain a system (Chaumun et al., 2000, Tahir et al., 2010). This suggests that it is reasonable to expect that coupling will be related to testability. Dynamic rather than static coupling has been selected for our investigation to address some shortcomings of the traditional static measures of coupling. For many years coupling has been measured statically, based on the limited structural properties of software (Zaidman and Demeyer, 2008). This misses the coupling at runtime between different components at different levels (classes, objects, packages, and so on), which should capture a more complete picture and so relate better to testability. This notion of measuring Dynamic Coupling is quite common in the emergent software engineering research literature. In our recent systematic mapping study of dynamic metrics, Dynamic Coupling was found to be the most widely investigated system characteristic used as a basis for dynamic analysis (Tahir and MacDonell, 2012).

For the purposes of this work the approach taken by (Arisholm et al., 2004) is followed, and Dynamic Coupling metrics that capture coupling at the object level are used. Two objects are coupled if at least one of them acts upon the other (Chidamber and Kemerer, 1994). The measure of coupling used here is based on runtime method invocations/calls: two classes, class A and class B, are said to be coupled if a method from class A (*caller*) invokes a method from class B (*callee*), or vice versa. Details of the specific metrics used to measure this form of coupling are provided in section 4.2.1.

## 3.2 Key Classes

The notion of a Key Class is introduced in this study as a new production code property to be measured and its relationship to class testability investigated. The meaning of Key Classes in this study is defined and its expected relationship to testability described.

OO systems are formed around groups of classes some of which are linked together. As software systems grow in size, so the number of classes used increases in these systems. To analyse and understand a program or a system, how it works and the potential for decay, it is important to know where to start and which aspects should be given priority. From a maintenance perspective, understanding the roles of classes and their relative importance to a system is essential. In this respect there are classes that could have more influence and play more prominent roles than others. This group of classes is referred to here as 'Key Classes'. We define a Key Class as a class that is executed frequently in the typical use profile of a system. Identifying these classes should inform the more effective planning of testing activities. One of the potential usages of these classes is in prioritizing testing activities – testers could usefully prioritize their work by focusing on testing these Key Classes first, alongside consideration of other factors such as risk and criticality information.

The concept of Key Classes is seen elsewhere in the literature, but has an important distinction in meaning and usage in this research. For example, in work of Zaidman and Demeyer (2008), classification as a Key Class is based on the level of coupling of a class. Therefore, Key Classes are those classes that are tightly coupled. In contrast, our definition is based on the *usage* of these classes: Key Classes are those classes that have high execution frequency at runtime. A metric used to measure Key Classes is explained in section 4.2.2.

The following section now describes and justifies the design of this study.

# 4 STUDY DESIGN

In this section we explain our research questions and the hypotheses that the work is aimed at testing. We also define the various metrics used in operational terms and our analysis procedures.

One of the key challenges faced when evaluating software products is the choice of appropriate measurements. Metric selection in this research has been determined in a "goal-oriented" manner using the GQM framework (Basili and Weiss, 1984) and its extension, the GQM/MEDEA framework (Briand et al., 2002). Our **goal** is to better understand what affects software testability, and our **objective** is to assess the presence and strength of the relationship between Dynamic Coupling and Key Classes on the one hand and code testability on the other. The specific **purpose** is to measure and ultimately predict class testability in OO systems. Our **viewpoint** is as software engineers, and more specifically, testers, maintainers and quality engineers. The targeted **environment** is Java-based open source systems.

## 4.1 Research Questions and Hypotheses

We investigate two factors that we contend are in principle related to system testability: Dynamic Coupling and Key Classes. For this purpose, we have two research questions to answer:

**RQ1**: Is Dynamic Coupling of a class significantly correlated with the internal class testability measures of its corresponding test class/unit?

**RQ2**: Are Key Classes significantly correlated with the internal class testability measures of their corresponding test classes/units?

The following two research hypotheses are investigated to answer the research questions:

**H0**: Dynamic Coupling has a significant correlation with class testability measures.

**H1**: Key Classes have a significant correlation with class testability measures.

The corresponding null hypotheses are:

**H2**: Dynamic Coupling has no significant correlation with class testability measures.

**H3**: Key Classes have no significant correlation with class testability measures.

## 4.2 Dynamic Measures

In section 3 we described the Dynamic Coupling and Key Classes testability concepts. In this section we define specific dynamic metrics that can be used to measure these testability concepts.

### 4.2.1 Dynamic Coupling Measures

As stated in subsection 3.1, Dynamic Coupling is intended to be measured in two forms - when a class is accessed by another class at runtime, and when a class accesses other classes at runtime (i.e., to account for both *callers* and *callees*). To measure these levels of coupling we select the previously defined *Import Coupling (IC)* and *Export Coupling (EC)* metrics (Arisholm et al., 2004). IC measures the number of method invocations *received* by a class (*callee*) from other classes (*callers*) in the system. EC measures the number of method invocations *sent* from a class (*caller*) to other classes (*callees*) in the system. Note that both metrics are collected based on method invocations/calls. More detailed explanations of these metrics are provided in Arisholm et al., (2004).

### 4.2.2 Key Classes Measure

The concept of Key Classes is explained in section 3.2. The goal here is to examine if those Key Classes (i.e., those classes with higher frequency of execution) have a significant relationship with class testability (as defined in the next subsection). We define the *Execution Frequency (EF)* dynamic metric to identify those Key Classes. EF for class $C$ counts the number of executions of methods within class $C$. Consider a class $C$, with methods $m_1, m_2,.....m_n$. Let $EF(m_i)$ be the number of executions of method $m$ of class $C$, then:

$$EF(C) = \sum_{i=1}^{n} EF(mi) \qquad (1)$$

*where n is the number of executed methods within class C*

## 4.3 Class Testability Measures

The testability of a class is considered here in relation to unit tests. In this work, we utilise two static metrics to measure unit test characteristics: Test Lines of Code (TLOC) and the Number of Test Cases (NTC). These metrics are motivated by the test suite metrics suggested by Bruntink and van Deursen (2006). TLOC, derived from the classic Lines of Code (LOC) metric, is a size measure that counts the total number of physical lines of code within a test class or classes. NTC is a test design metric that counts the total number of test cases in a test class.

Our hypotheses thus reflect an expectation that the Dynamic Coupling and Key Classes of production code classes are related to the size and scope of their associated test classes.

## 4.4 Testing the Relationships

As we are interested in the potential associations between variables, a statistical test of correlation is used in the evaluation of our hypotheses. After collecting our metrics data we first apply the Shapiro-Wilk (S-W) test to check the normality of each data distribution. This is necessary as selection of the relevant correlation test should be informed by the nature of the distributions, being normal or non-normal. The S-W test is a particularly appropriate one to use here given the size of our data sets (as detailed in the next section). The null hypothesis for the S-W test is that data is normally distributed. Our data collection methods are explained in more detail in the following section.

# 5   DATA COLLECTION

The collection of dynamic metrics data can be accomplished in various ways. The most common (and most accurate) way is to collect the data by obtaining trace information using dynamic analysis techniques during software execution. Such an approach is taken in this study and is implemented by collecting metrics using the *AspectJ*[1] framework, a well-established Java implementation of Aspect Oriented Programming (AOP). Previous works (including those of Cazzola and Marchetto (2008), Adams et al. (2009) and Tahir et al. (2010)) have shown that AOP is an efficient and practical approach for the objective collection of dynamic metrics data, as it can enable full runtime automatic source-code instrumentation to be performed.

Testability metrics data, including LOC, TLOC, and Number of Classes (NOC), are collected using the *CodePro Analytix*[2] tool. The values of these metrics were later checked and verified using the *Eclipse Metrics Plugin*[3]. Values for the NTC metric are collected from the *JUnit*[4] framework and these values were verified manually by the first author.

We used the two different traceability techniques suggested by Rompaey and Demeyer (2009) to identify unit test classes and link them to their corresponding production classes. First, we used the *Naming Convention* technique to link test classes to production classes following their names. It has been widely suggested (for instance, in the JUnit documentation) that a test class should be named after the corresponding class(es) that it tests, by adding "Test" to the original class name. Second, we used a *Static Call Graph* technique, which inspects method invocations in the test case. The latter process was carried out manually by the first author. The effectiveness of the Naming Convention technique is reliant on developers' efforts in conforming to a coding standard, whereas the Static Call Graph approach reveals direct references to production classes in the test classes.

It is important to note here that we only consider core system code: only production classes that are developed as a part of the system are assessed. Additional classes (including those in jar files) are excluded from the measurement process. These files are generally not part of the core system under development and any dependencies could negatively influence the results of the measurement process.

## 5.2 5.1 Case Studies

To consider the potential relationships between class testability and the selected dynamic metrics we selected three different open source systems to be used in our experiments. The selection of these systems was conducted with the goal of examining applications of reasonable size, with some degree of complexity, and easily accessible source code. The main criteria for selecting the applications are: 1) each application should be fully open source i.e., source code (for both production code and test code) is publicly available; 2) each application must be written in Java, as we are using the JUnit and AspectJ frameworks, which are both written for Java; 3) each application should come with test suites; and 4) each application should comprise at least 25 test classes.

The systems selected for our experiments are: *JabRef*[5], *Dependency Finder*[6] *and MOEA*[7]. Brief descriptions of the selected systems are shown in Table 1. Table 2 reports particular characteristics and size information of both the production and test code of the three systems.

Table 1: Brief descriptions of the selected systems.

| System | Description |
|---|---|
| *JabRef* | A cross-platform bibliography tool that provides GUI-based reference management support for the BibTeX file format – a LaTeX based referencing format. |
| *Dependency Finder* | An analyser tool that extracts dependencies, develops dependency graphs and provides basic OO metric information for Java compiled code. |
| *MOEA* | A Java-based framework oriented to the development and experimentation of multi-objective evolutionary and optimization algorithms. |

The size classification used in Table 2 is adapted from the work of Zhao and Elbaum (2000), where application size is categorised into bands based on the number of kiloLOC (KLOC): small (fewer than 1 KLOC), medium (1-10 KLOC), large (10-100 KLOC) and extra-large (more than 100 KLOC).

## 5.3 Execution Scenarios

In order to arrive at dynamic metrics values that are associated with typical, genuine use of a system the selected execution scenarios must be representative of such use. Our goal is to mimic 'actual' system behaviour, as this will enhance the utility of our results. The scenarios are therefore designed to use the key system features, based on the available documentation and user manuals for the selected systems, as well as our prior knowledge of these systems. Further information on the selected execution scenario for each system now follows. Note that all three systems have GUI access, and the developed scenarios assume use via the GUI.

*JabRef*: the tool is used to generate and store a list of references from an original research report. We included all reference types supported by the tool (e.g., journal articles, conference proceedings, reports, standards). Reports were then extracted using all available formats (including XML, SQL and CSV). References were managed using all the provided features. All additional plugins provided at the tool's website were added and used during this execution.

*Dependency Finder*: this scenario involves using the tool to analyse the source code of four medium-large sized systems one after another, namely, FindBugs,

JMeter, Ant and Colossus. We computed dependencies (dependency graphs) and OO metrics at all layers (i.e., packages, classes, features). Analysis reports on all four systems were extracted and saved individually.

*MOEA*: *MOEA* has a GUI diagnostic tool that provides access to a set of 6 algorithms, 57 test problems and search operators. We used this diagnostic tool to apply those different algorithms on the predefined problems. We applied each of these algorithms at least once on each problem. We displayed metrics and performance indicators for all results provided by those different problems and algorithms. Statistical results of these multiple runs were displayed at the end of the analysis.

## 6 RESULTS

On applying the S-W test to our data for all three systems the evidence led us to reject the null hypothesis regarding their distribution, and so we accepted that the data were not normally distributed (see Figures 1-3 for illustration). We therefore decided to use Kendall's tau ($\tau$) rank coefficient test. Kendall's tau is a rank-based non-parametric statistical test that measures the association between two measured quantities. In our work Kendall's tau is calculated to assess the degree of association between each dynamic metric of the production code (i.e., IC, EC and EF) and the class testability metrics, defined in sections 4.2 and 4.3 respectively.

We used the classification of Cohen (1988) to interpret the degree of association between variables. The value of $\tau$ indicates the association between two ranked variables, and it ranges from -1 (perfect negative correlation) to +1 (perfect positive correlation). We interpret that variables are when $\tau = 0$, that there is a low direct association when $0 < \tau \leq 0.29$, a medium direct association when $0.3 \leq \tau \leq 0.59$, and a strong direct association when $0.6 \leq \tau \leq 1$. This interpretation also applies to negative correlations, but the association is inverse rather than direct (Daniel, 2000). The $p$ value represents the statistical significance of the relationship. We consider an association to be statistically significant where $p \leq 0.05$.

The number of observations considered in each test varies in accordance with the systems' execution scenarios described in subsection 5.2. Observation points, in fact, represent the number of tested classes that were traversed in the execution (viz. classes that have corresponding tests and were captured during the execution by any of the dynamic metrics used). The number of observations for *JabRef* is 26, 80 for *Dependency Finder* and 76 for *MOEA*.

Table 2: Characteristics of the selected systems.

| System | Version | KLOC | Size | NOC | # JUnit classes | NTC | Test KLOC |
|--------|---------|------|------|-----|-----------------|-----|-----------|
| *JabRef* | 2.9.2 | 84.717 | Medium | 616 | 55 | 237 | 5.392 |
| *Dependency Finder* | 1.2.1 beta4 | 26.231 | Medium | 416 | 258 | 2,003 | 32.095 |
| *MOEA* | 1.17 | 24.307 | Medium | 438 | 280 | 1,163 | 16.694 |

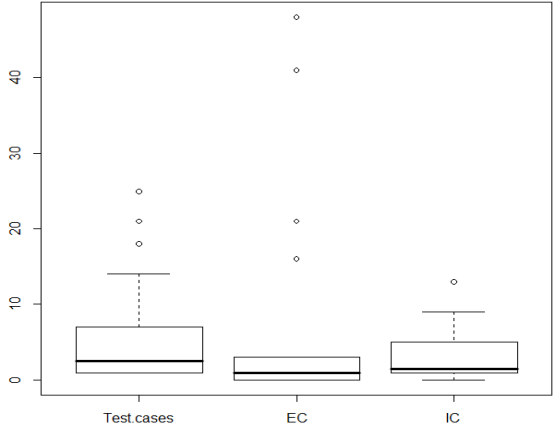

Figure 1: Data distribution boxplots for the JabRef system.

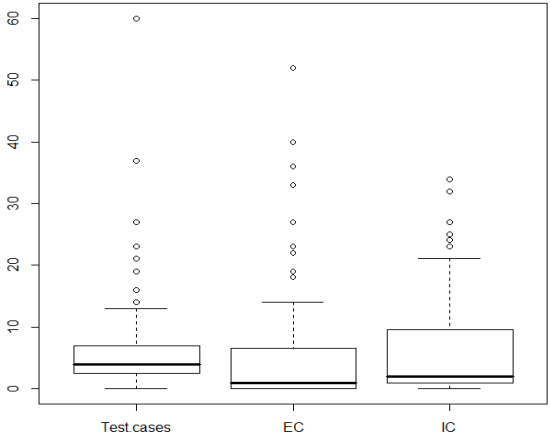

Figure 2: Data distribution boxplots for the Dependency Finder system.

Table 3 shows the Kendall's tau results for the two dynamic coupling metrics against the test suite metrics. Corresponding results for the execution frequency (EF) metric against the test suite metrics are presented in Table 4.

For dynamic coupling, we see (Table 3) a mix of results from the collected metrics. EC is observed to have a significant relationship with the TLOC metric in two of the three systems. These relationships vary from low direct (in the case of *MOEA*) to medium direct (in the *Dependency Finder* case). However, a similar significant correlation between EC and NTC is only evident for the *Dependency Finder* system (a medium direct association).

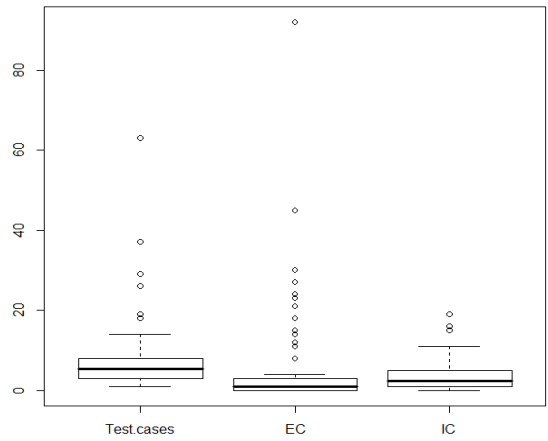

Figure 3: Data distribution boxplots for the MOEA system.

In terms of relationships with the NTC metric (Table 4), a low direct association between IC and NTC is evident in the case of *JabRef*. Analysis of *Dependency Finder* reveals a significant medium direct association between these metrics. A low inverse association between IC and NTC is evident for the *MOEA* system.

Positive and significant associations were found between EF and the test suite metrics for two of the three systems (the exception being the *MOEA* system). We found a significant, medium direct association between EF and TLOC and between EF and NTC in the case of *JabRef*. In *Dependency Finder*, low direct associations between EF and both TLOC and NTC were revealed.

Table 3: Dynamic coupling correlation results.

| *Systems* | *Metrics* | *TLOC* | | *NTC* | |
|---|---|---|---|---|---|
| | | τ | p | τ | p |
| *JabRef* | EC | .292 | .054 | .291 | .068 |
| | IC | .193 | .193 | **.148** | **.041** |
| *Dependency Finder* | EC | **.389** | **.000** | .319 | .000 |
| | IC | **.388** | **.000** | **.251** | **.003** |
| *MOEA* | EC | **.230** | **.008** | .093 | .300 |
| | IC | -.055 | .504 | **-.190** | **.027** |

Table 4: Execution Frequency (EF) correlation results.

| *Systems* | *Metrics* | *TLOC* | | *NTC* | |
|---|---|---|---|---|---|
| | | τ | p | τ | p |
| *JabRef* | EF | **.344** | **.016** | **.306** | **.041** |
| *Dependency Finder* | EF | **.216** | **.005** | **.158** | **.048** |
| *MOEA* | EF | .005 | .953 | -.074 | .366 |

# 7 DISCUSSION

Based on our analysis we accept **H0** and reject **H2**; that is, we note evidence of a significant association between dynamic coupling (either EC or IC) and the two test suite metrics for all three systems analysed here. As we also found EF to be significantly associated with the test suite metrics for two of the three systems considered we also accept **H1** and reject **H3** on the balance of evidence.

An additional test of relevance in this study is to consider whether our dynamic testability metrics are themselves related, as this may indicate that only a subset of these metrics needs to be collected. We therefore performed further correlation analysis to investigate this.

Our results indicate that the Dynamic Coupling metrics are correlated with EF (Table 5) to varying degrees for the three systems investigated. High direct and medium direct associations between one or both of the two Dynamic Coupling metrics (i.e., IC and EC) and the EF metric are evident for all three systems.

Table 5: Correlation results between coupling and EF dynamic metrics.

| | *Metrics* | **IC** | | **EC** | |
|---|---|---|---|---|---|
| | | τ | p | τ | p |
| **EF** | *JabRef* | .194 | .198 | **.691** | **.000** |
| | *Dependency Finder* | **.415** | **.000** | **.376** | **.000** |
| | *MOEA* | **.221** | **.008** | **.304** | **.000** |

In summary, we found EC to have a significant correlation with TLOC, where IC was significantly associated with NTC. We interpret this to indicate that Dynamic Coupling, in some form, has a significant correlation with test suite metrics. We draw a similar inference regarding Key Classes; this property is also significantly associated with our test suite metrics. Additionally, we found the two dynamic testability concepts studied here, i.e., Dynamic Coupling and Key Classes, to be themselves significantly correlated.

In revisiting our research questions, we found Dynamic Coupling to have a significant (although not strong) direct association with testability metrics (RQ1). A more significant correlation was found between Key Classes (i.e., frequently executed classes) and class testability metrics. By answering RQ1 and RQ2, we suggest that Dynamic Coupling and Key Classes can act, to some extent, as complementary indicators of class testability. We contend here that a tightly coupled or frequently executed class would need a large corresponding test class (i.e., higher numbers of TLOC and NTC). This expectation has been found to be evidenced in at least two of the three systems examined.

# 8 THREATS TO VALIDITY

We acknowledge a number of threats that could affect the validity of our results.

**- Limited number and form of systems**: The results discussed here are derived from the analysis of three open source systems. The consideration of a larger number of systems, perhaps also including closed-source systems, could enable further evaluation of the associations revealed in this study.

**- Execution scenarios**: All our execution scenarios were designed to mimic as closely as possible 'actual' system behaviour, based on the available system documentation and, in particular, indications of each system's key features. We acknowledge, however, that the selected scenarios might not be fully representative of the typical uses of the systems. Analysing data collected based on different scenarios might give different results. This is a very common threat in most dynamic analysis research. However, we tried to mitigate this threat by carefully checking user manuals and other documentation of each of the examined systems and deriving the chosen scenarios from these sources. Most listed features were visited (at least once) during the execution. We are planning to examine more scenarios in the future and compare the results from these different scenarios.

**- Testing information:** Only available test information was used. We did not collect or have access to any information regarding the testing strategy of the three systems. Test strategy and criteria information could be very useful if combined with the test metrics, given that test criteria can inform testing decisions, and the number of test cases designed is highly influenced by the implemented test strategy.

**- Test class selection:** We only considered production classes that have corresponding test classes, which may lead to a selection bias. Classes that are extremely difficult to test, or are considered too simple, might have zero associated test classes. Such production classes are not considered in our analyses. Due to their availability, we only included classes that had associated JUnit test classes, and ignored all others.

# 9  CONCLUSIONS AND FUTURE WORK

In this work we set out to investigate the presence and significance of any associations between two runtime code properties, namely Dynamic Coupling and Key Classes, and the internal testability of classes in three open source OO systems. Testability was measured based on the systems' production classes and their associated unit tests. Two different metrics were used to measure internal class testability, namely TLOC and NTC. As we were interested in the relationships between system characteristics at runtime, Dynamic Coupling and Key Classes were measured using dynamic software metrics collected via AOP. Results were then analysed statistically using the Kendall's tau coefficient test to study the associations.

The resulting evidence indicates that there is a significant association between Dynamic Coupling and internal class testability. We found that Dynamic Coupling metrics, and especially the export coupling metric (EC), have a significant direct association with TLOC. A less significant association was found between dynamic import coupling (IC) and NTC. Similarly, Key Classes are also shown to be significantly associated with our test suite metrics in two of the three systems examined.

The findings of this work contribute to our understanding of the nature of the relationships between characteristics of production and test code. The use of dynamic measures can provide a level of insight that is not available using static metrics alone. These relationships can act as an indicator for internal class level testability, and should be of help in informing maintenance and reengineering tasks.

Several future research directions are suggested by the outcomes of this research. This work can be extended by examining a wider range of systems (such as closed-source systems) to enable further evaluation of the findings. Another research direction would be to investigate whether Dynamic Coupling and Key Class information can be used together to predict the size and structure of test classes. Predicting class-level testability should improve the early estimation and assessment of the effort needed in testing activities. This work could also be extended to an investigation of the association between other source code factors and testability using runtime information. It would also be potentially beneficial to incorporate the current information about class testability with other testing information such as test coverage and test strategy.